# Waveforms for sub-THz 6G: Design Guidelines

Muris Sarajlić[*], Nuutti Tervo[†], Aarno Pärssinen[†], Le Hang Nguyen[§], Hardy Halbauer[§], Kilian Roth[+], Vaidyanathan Kumar[‖], Tommy Svensson[‖], Ahmad Nimr[††], Stephan Zeitz[††], Meik Dörpinghaus[††], Gerhard Fettweis[††]

[*]Ericsson AB, Sweden, [†]University of Oulu, Finland, [§]Nokia Bell Labs, Germany, [+]Intel Deutschland, Germany, [‖]Chalmers University of Technology, Sweden, [††]Vodafone Chair Mobile Communication Systems, TU Dresden, Germany

muris.sarajlic@ericsson.com

*Abstract*—The projected sub-THz (100 – 300 GHz) part of the upcoming 6G standard will require a careful design of the waveform and choice of slot structure. Not only that the design of the physical layer for 6G will be driven by ambitious system performance requirements, but also hardware limitations, specific to sub-THz frequencies, pose a fundamental design constraint for the waveform. In this contribution, general guidelines for the waveform design are given, together with a non-exhaustive list of exemplary waveforms that can be used to meet the design requirements.

## I. Introduction

What will the sixth generation (6G) of mobile communications provide and entail? Pre-standardization research activities are currently ongoing that aim at providing an answer to this question. The vision commonly held is that 6G will provide a technological platform enabling a unification of physical, digital and human worlds [1][2]. Technical features such as incorporating AI into the network and use of sensing will help fulfill a set of (primarily societal) goals, namely, providing connectivity for everyone and anywhere; supporting sustainable development; and achieving all this in a trustworthy manner [2][3].

Along with the target and vision of what 6G will enable, a picture is being formed of the underlying technical enablers together with performance requirements. Requirements and enablers, naturally, vary from one use case to another. Some of the envisioned use cases, such as digital twins and holographic communication, will require extreme peak throughputs on the order of 100 Gbps [4]-[6]. Additionally, some use cases require ultra-low latencies, down to 0.1 ms of user plane latency [4][6].

It is generally assumed that one of the key moves made towards meeting and reconciling the ambitious and sometimes conflicting 6G performance requirements will be to use essentially *all of the known radio frequency* (RF) *spectrum*, in a flexible way [3]: from sub-6 GHz and coexistence with existing allocations there, via exploiting the 7 – 15 GHz bands not previously used for mobile communications, to a more efficient use of the currently under-exploited lower millimeter wave (mmW) bands (24 – 52 GHz) that may eventually also incorporate bands towards 100 GHz, and finally, exploiting the abundant supply of spectrum that lies in the 100 – 300 GHz frequency range, here and elsewhere referred to as the sub-THz bands.

Bandwidth found in vast amounts in the sub-THz frequency range is likely to be the primary enabler of the >100 Gbps peak throughput as referenced above. (In total, 50.7 GHz and 46.5 GHz of non-contiguous bandwidth is allocated for fixed or mobile services in the 100 – 200 GHz and 200 – 300 GHz frequency ranges, respectively [7][8]). In addition to enabling extreme per-user throughputs, sub-THz in 6G could e.g. be used for offloading some of the mobile traffic from lower bands or for providing high-throughput fixed wireless access service. Moreover, localization and sensing (envisioned in 6G as integrated with communication) will require several GHz of bandwidth for supporting some of the high-resolution use cases [5], and some or all of that bandwidth could also come from sub-THz bands.

In this contribution, an overview is given of design guidelines for waveform design at sub-THz frequencies as influenced by the known and projected hardware and system limitations. Additionally, examples of waveforms fulfilling the design guidelines are provided.

## II. Hardware limitations and system design constraints

Operation at sub-THz frequencies is limited by a number of fundamental physical constraints exhibiting unfavorable scaling with carrier frequency $f_c$. Free-space propagation loss scales as $f_c^2$ due to shrinking antenna aperture and compensating that by adding more antenna elements leads easily to a very large parallelism at upper mmW region. Peak output power of integrated power amplifiers (PAs) reduces with frequency due to shrinking circuit size and scales approximately as $f_c^{-3}$ beyond 100 GHz [9]; likewise, PA efficiency exhibits a rapid decrease with $f_c$. The bad efficiency combined with very small circuit footprint poses serious thermal challenges [16]. Hence, despite of the low power per PA, efficiency under backoff is extremely important figure of merit to improve. Large number of parallel PAs also challenge the digital linearization techniques used to linearize the transmitter while making PA behavior dependent on beamforming [17]. Phase noise (PN) power increases as $f_c^2$ due to frequency multiplications inherent to carrier signal synthesis. Also, noise figure of the receiver increases exponentially when frequency approaches limits of the specific semiconductor technology.

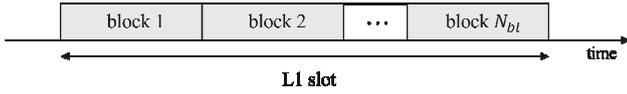

Figure 1: time-domain structure of L1 slot

Wideband operation on bandwidths of several GHz or tens of GHz, as projected for sub-THz, also presents a range of design difficulties. One of those is the increased complexity and power consumption of digital signal processing circuitry, as well as increased power consumption of the analog-to-digital/digital-to-analog converters (ADC/DAC): the data in [10] indicates that the power consumption of the ADC/DACs scales as the square of the sampling rate for sampling rates larger than 500 MHz and when the ADC/DAC dynamic range is kept constant. Wideband operation also increases the impact of thermal noise in receiver and makes the flat part of the phase noise spectra to have impact on the overall performance [18]. Using very wide bandwidth also makes many analog baseband nonidealities such as IQ imbalance very frequency selective. Furthermore, wideband operation makes PA memory effects more severe.

The design of wireless communication systems operating at sub-THz needs to counter the detrimental effects of hardware limitations including ones listed above and even more as in [11]. Careful hardware-aware waveform design may improve the transmit power, robustness to impairments and improve energy efficiency, as will be elaborated shortly.

In addition to hardware limitations listed above, projected 6G system requirements will also have leverage on waveform design. Besides required throughput and reliability, it is also user-plane latency that will have impact on the waveform structure, and in particular the structure of the physical layer slot. This impact is analyzed in the following.

### III. SLOT STRUCTURE AND PROCESSING TIME AS FUNCTION OF THE LATENCY REQUIREMENT

Some of the more latency-sensitive use cases for 6G (e.g. robot-assisted surgery, digital twins for manufacturing) will require <1 ms application layer latency [4], achieving which would require radio access network (RAN) latency significantly lower than 1 ms. In Hexa-X, RAN user-plane latency requirement for the most extreme use-cases is set to <100 µs [12].

Table 1: Example sampling rates, block and slot durations and bandwidth

| SCS [MHz] | sampling time, $T_s$ [ns] | $T_{block}$ = (2048+144) $T_s$ [µs] | $T_{slot}$ = $2T_{block}$ [µs] | $T_{slot}$ = $14T_{block}$ [µs] | BW = $0.6/T_s$ [GHz] |
|---|---|---|---|---|---|
| 480 | 1.02 | 2.23 | 4.46 | 31.2 | 0.59 |
| 960 | 0.5 | 1.11 | 2.23 | 15.6 | 1.18 |
| 1920 | 0.25 | 0.56 | 1.11 | 7.8 | 2.36 |
| 3840 | 0.13 | 0.28 | 0.56 | 3.9 | 4.72 |

The analysis presented in this section is applicable to any waveform and physical layer structure. In this way, it is possible to demonstrate what kind of changes to the current design of 5G NR physical layer and of the related signal processing are necessary.

The definition of RAN user-plane latency used here is adopted from 3GPP as *the time it takes to successfully deliver an application layer packet from radio protocol layer 2/3 ingress point to the radio protocol layer 2/3 egress point* [13]. Packet retransmissions (in form of HARQ) are handled by layer 2 (MAC) and thus calculation of RAN latency needs to take the retransmissions into account

A very generic time-domain frame structure is adopted, illustrated in Figure 1, where $N_{bl}$ blocks are concatenated in time to form a L1 slot. What the block is depends on the waveform used; e.g. if the waveform is OFDM, the block is an OFDM symbol plus a cyclic prefix. To keep the analysis comparable to 5G NR L1 structure, sampling rates are connected to OFDM subcarrier spacing (SCS). The time structure for several sampling rates/subcarrier spacings of importance is given in Table 1. Two different slot sizes (short, with 2 blocks and long, with 14 blocks) are selected, and the bandwidth used for transmissions is assumed to occupy 60% of the theoretical maximum bandwidth, thus allowing for frequency guard bands.

For the processing times, it is assumed that

$$T_{proc,\,tx} = T_{proc,\,rx} = \alpha 80 \text{ µs} \qquad (4)$$

and

$$T_{proc,\,tx} = T_{proc,\,rx} = \alpha 98.2 \text{ µs}, \qquad (5)$$

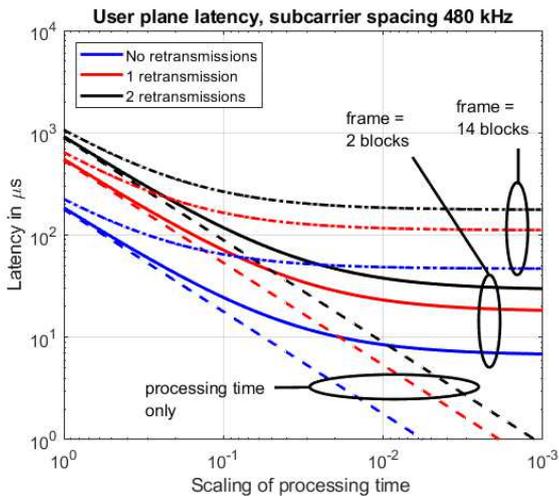

Figure 2: User-plane latency in a "low-mmW" scenario: system BW ~ 0.6 GHz. Dashed: contribution of processing time to overall latency

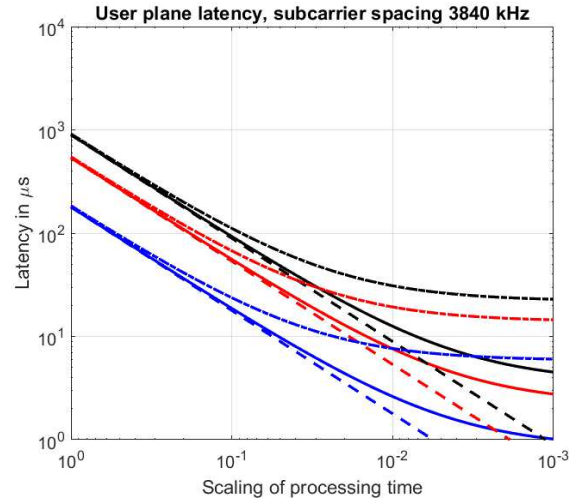

Figure 3: User-plane latency in a "sub-THz" scenario: system BW ~ 4.7 GHz

where scaling factor α ∈ [0, 1] and the numerical values 98.2 µs and 80 µs are state-of-the-art UE and BS processing times per slot as recognized by 3GPP [14][15]. In other words, the analysis examines what kind of influence the downscaling of currently available slot processing time has on latency.

The results of the parametric analysis of latency as per the discussion above are shown in Figure 2 and Figure 3. The "waveform" in the scenario in Figure 2 has the same Nyquist sampling rate as an OFDM waveform with SCS of 480 kHz, which would correspond to a system bandwidth of about 0.6 GHz as suggested in Table 1. This scenario is therefore applicable to current 5G NR deployments, or a 6G deployment on a lower carrier frequency, e.g. the 24 – 52 GHz band. The results show that the 100 µs latency requirement cannot be met without significant reduction of slot processing time. Reducing the slot size from 14 to 2 blocks/OFDM symbols also helps reduce latency, in particular when there are retransmissions.

For the scenario in Figure 3, where sampling rate corresponds to a bandwidth of 4.7 GHz and is thus of relevance to sub-THz deployments, block and slot durations are short and thus their contribution to overall latency is minor; most of the latency comes from the processing time. We can generally say that the **slot processing time needs to go down to 10 – 50 µs in order for the user-plane latency target of 100 µs to be achieved**. The part of the slot processing time proportional to the amount of data contained in the slot will be reduced by using shorter slots (e.g. containing 1, 2 or 4 blocks/OFDM symbols). Additionally, using waveforms and channel codes enabling parallelized Tx/Rx processing will help reduce processing times further. Furthermore, iterative receiver processing algorithms should be carefully designed so as not to break the processing time budget.

IV. GUIDELINES ON HARDWARE-CONSTRAINED WAVEFORM DESIGN

Regarding the hardware-related challenges outlined in Section II, it is becoming clear that alternatives to OFDM (which has been the waveform of choice in 4G and 5G) need to be sought for the sub-THz part of 6G. Robustness of OFDM to uncompensated phase noise is poor due to phase noise causing intercarrier leakage, and the envelope variations of OFDM are high, directly translating to larger output power backoff of the PA and thus decrease of coverage. Moreover, as analog and hybrid beamforming with very narrow beams will be prevalent at sub-THz, the need for very flexible frequency scheduling of users – a major advantage of using OFDM – will be less pronounced in most use cases due to only a few users simultaneously being served by the same beam. All this is not to say that OFDM would still not be applicable or even necessary in some scenarios even at sub-THz; but in most cases, other waveforms might be better suited for the task. Also, it should be noted that OFDM will likely still be highly relevant for 6G in lower frequency bands.

A general observation to make here is that the physical layer of 6G will likely be based on several waveforms, enabling a certain degree of dynamic switching between the waveforms depending on the beamforming used (analog/digital), operating band (e.g. 7-15 GHz vs. mmWave vs. sub-THz), whether coverage/throughput/energy efficiency is being maximized at a certain operating point, if sensing is to be used etc. To this end, it would be beneficial if **6G waveforms would be chosen such that they fit one common block and frame structure** and also support a certain degree of hardware reuse (so that switching between waveforms means simply turning one or few DSP blocks on or off). A related concept, denoted as "Gearbox PHY" in [23], aims at maximizing energy efficiency by providing multiple radio design options (gears) in terms of hardware and waveforms that support different spectral efficiency. Depending on the traffic requirements and the channel conditions, the radio can be switched to the suitable "gear".

Another important consideration is that the choice of waveform will have impact on the link budget both at the transmitter - most importantly, it will set the PA backoff - and at the receiver, where it will determine the sensitivity. For most purposes, evaluations of waveform suitability therefore **must take into account the impact of waveform on the entire link, end-to-end**.

A. *Power amplifier constraints*

As discussed above, PA's low efficiency properties at frequencies above 100 GHz dictate the use of waveforms with low envelope variation. DFT-spread OFDM (DFT-s-OFDM, also sometimes referred to as SC-FDMA) is an alternative waveform to OFDM offering reduced envelope variations, adopted for the UL of 4G and 5G NR. Single carrier (SC) presents another viable alternative to multi-carrier transmission. It was shown in [19][21] that Single Carrier with Frequency Domain Equalization (SC-FDE), using QAM modulation and common root raised cosine as pulse shaping filter, possesses the lowest PAPR level compared to OFDM and DFT-s-OFDM without frequency domain filtering [19]. Thanks to the simple transmit structure of SC-FDE design, further PAPR reduction can be achieved in modifying the modulation alphabet and properly designing the band limited

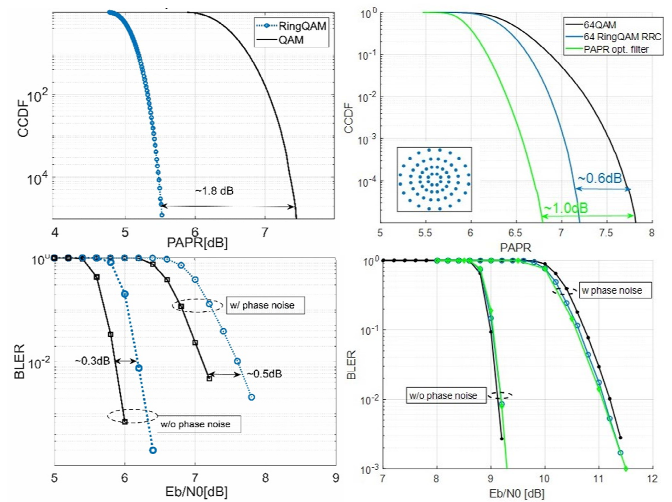

Figure 4: PAPR CCDF and BLER for 16-APSK (left) and 64-APSK (right)

Table 2: SNR in dB at which BLER=0.1 is achieved.

| OPBO in dB | DFT-s-OFDM QPSK | SC-FDE QPSK | DFT-s-OFDM 16-QAM | SC-FDE 16-QAM | DFT-s-OFDM 64-QAM | SC-FDE 64-QAM |
|---|---|---|---|---|---|---|
| 3 | NA | 4.12 | NA | NA | NA | NA |
| 4 | 7.00 | 2.39 | NA | NA | NA | NA |
| 5 | 3.17 | 1.92 | NA | 9.11 | NA | NA |
| 6 | 1.96 | 1.70 | 10.7 | 8.01 | NA | 14.5 |
| 7 | 1.55 | 1.55 | 8.20 | 7.72 | 18.1 | 13.4 |
| 8 | 1.36 | 1.46 | 7.67 | 7.57 | 13.8 | 13.0 |

transmit filter.

For OFDM, the impact on PAPR from choosing alternative QAM schemes is almost negligible. However, for single carrier signals the impact of the QAM shape is relevant, as also recognized e.g., by the IEEE DVB-S2 standard for satellite broadcasting [20] which utilizes single-carrier as transmission waveform and Amplitude Phase Shift Keying (APSK) constellation patterns as modulation scheme. A study in [21] showed that for the selected exemplary 16- and 64-APSK constellation a PAPR reduction of 1.8dB and 0.6dB can be achieved at a moderate loss of 0.3dB and 0dB respectively in the BLER performance, which is caused by the reduced Euclidean distance of the constellation points compared to the standard QAM constellation (Figure 4).

Another means for SC-FDE waveforms to further constrain the envelope variation is the optimization of the transmit pulse shaping filter. Commonly, a root raised cosine filter with a certain roll-off is used to limit the bandwidth and shape the transmit pulses of a SC transmission. The roll-off factor defines the excess bandwidth and the slope of the resulting pulse. Explicitly designing the transmit pulse shaping filter (in combination with the corresponding adaptation of the receive filter) to minimize the peak amplitude of the transmit signal allows to further reduce the PAPR level. An example for such a PAPR aware filter (green curves) is shown on the right-hand side of Figure 4 where an additional 0.4 dB reduction in PAPR can be harvested.

As an additional result, the results of link level simulations shown in Table 2 compare the performance of DFT-s-OFDM and SC-FDE at different Output Power Back Offs (OPBOs). Most system parameters used for the evaluations are ones developed for 3GPP NR adapted to support a system operating in the carrier frequency of 140 GHz. The used PA models are based on the GaN PA presented in [12]. Based on this model the results show that with SC-FDE a 1-2 dB lower OPBO can be used for achieving the same BLER at a 0.75 coding rate.

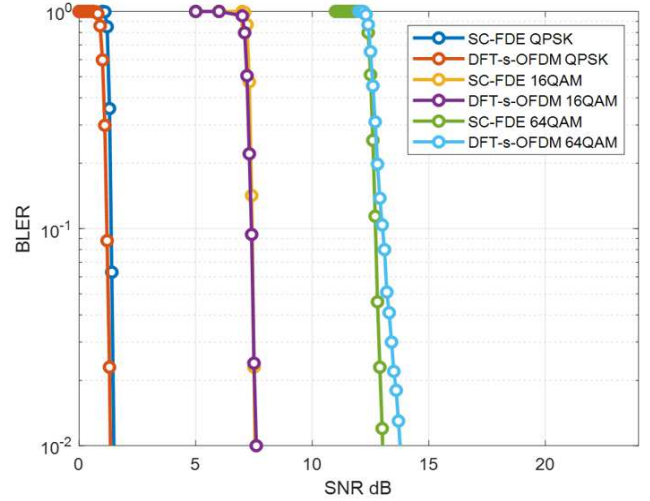

Figure 6: performance comparison of SC-FDE and DFT-s-OFDM with PN

In summary, SC-FDE transmissions can properly be designed to be a PA friendly waveform. This promotes it to a potential candidate for transmissions in scenarios in which PA output power is a limiting factor and power/energy efficiency is of cardinal relevance.

Another potential approach to mitigate the effects of non-linear PAs is to constrain the envelope variations by non-linear correlated mapping of the modulation symbols using carefully chosen waveform segments in a trellis-based pre-coder before mapping to a DFT-s-OFDM modulator. Such a scheme has been proposed in [28], using a sub-sampled continuous phase modulation (CPM) encoder as precoder, and interleaved frequency division multiple access subcarrier mapping of the samples. The scheme can provide a very low PAPR signal with very low side lobes, provided the CPM parameters, over-sampling factor and subcarrier mapping are suitably selected [28]. A drawback is the decreased receiver sensitivity by constraining the signal space, so it is interesting to investigate the overall performance taking not only the PA efficiency into account, but also the receiver performance. Table 3 shows the results of such a study, i.e. the cases in which CPM-SC-FDMA outperforms convolutionally encoded QPSK-SC-FDMA in terms of coverage, under the same target on spectral efficiency and bit error rate, as well as encoder memory. The results have been obtained using a real PA using RF WebLab

Table 3: Coverage of CPM-SC-FDMA vs QPSK-FDMA, one information bit per symbol, and target BER=0.0001

| Gate Voltage (V) | Best Coverage |
|---|---|
| -2.3 | QPSK-DFT-s-OFDM |
| -2.5 | CPM- DFT-s-OFDM with modulation index $h = 1/2$ |
| -2.75 | CPM- DFT-s-OFDM with modulation index $h = 1/2$ |
| -3 | CPM- DFT-s-OFDM with modulation index $h = 1/2$ |
| -3.3 | CPM- DFT-s-OFDM with modulation index $h = 1/2$ |
| -3.75 | CPM- DFT-s-OFDM with modulation index $h = 1/2$ |
| -4 | QPSK- DFT-s-OFDM |
| -4.5 | CPM- DFT-s-OFDM with modulation index $h = 1/2$ |

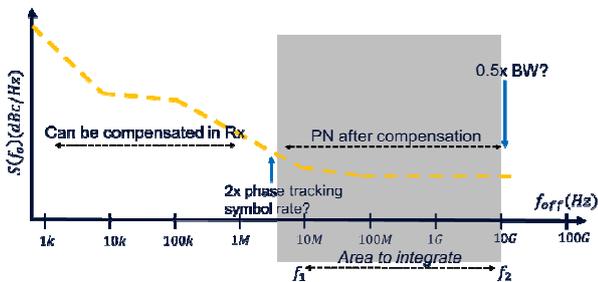

Figure 5: overview of the generic waveform perspective on a typical PN spectrum

[29], in which the gate voltage is varied to generate various degrees of PA non-linearity and overall efficiency of the PA with the modulated signals.

*B. Phase noise constraints*

The effect of PN from a local oscillator on the waveform can be observed as a combination of two separate processes: a "slow" one, and a "fast" one. The effects of "slow" PN can typically be compensated at the receiver; compensating the effects of "fast" PN is harder and some waveforms may inherently possess a comparatively higher level of robustness towards "fast" PN.

To clarify this matter further, Figure 5 shows a typical single sideband (SSB) PN PSD [21] of a phased locked loop (PLL) output. The overall effective phase noise power is an integral of PN PSD from the "lowest offset frequency that matters" up to the bandwidth of the signal. From the waveform perspective, the high sampling rate has pros and cons for the waveform phase noise tolerance. The good thing is that the slow phase variations that contain long memory can be effectively compensated with reasonable overhead with help of time domain reference symbols. This shifts the lower bound of the integral in Figure 5 to the right. On the other hand, fast variations appearing in the flat part of the phase noise spectra cannot be effectively compensated in the Rx [21] and remain as the dominant phase noise power after compensation due to the wide signal bandwidth. The phase noise scaling with frequency is severe: -145 dBc/Hz phase noise floor with 10 GHz PLL output frequency, would mean around -121 dBc/Hz at 150 GHz, and -115 dBc/Hz at 300 GHz. This spread over 10 GHz of signal bandwidth makes roughly 21 dB SNR at 150 GHz and 15 dB at 300 GHz.

The link levels simulations shown in Figure 6 compare the performance of DFT-s-OFDM and SC-FDE using the phase noise model developed in [21]. Most system parameters used for the evaluations are ones developed for 3GPP NR adapted to support a system operating in the carrier frequency of 140 GHz, using a coding rate of 0.75. The results show that using a per block compensation of the phase noise for both systems leads to a slightly better result for SC-FDE for 64-QAM. It needs to be noted that phase noise performance that can be expected in a future, consumer grade system at this frequency is still unclear.

*C. Ultra-wide bandwidth constraints*

As noted in Section II, ADC/DAC power consumption is a problem for systems with several GHz or tens of GHz of bandwidth. One solution to this problem is reducing the number of quantization bits; this will, naturally, have a negative impact on decoding error probability. This negative effect could be reversed by finding a waveform that is robust to coarse quantization. For example, it has been shown in [19][21] that SC-FDE is more robust against quantization noise than DFT-s-OFDM without FD filtering. In terms of ADC power consumption, [19][21] showed that the power consumption deploying SC-FDE waveform can be of a factor of 2 lower compared to the case of applying DFT-S-OFDM, considering 64-QAM transmissions. This reduction in the power consumption linearly scales up with the number of RF chains and becomes dominant for architectures with a larger number of RF chains at the receive side [22].

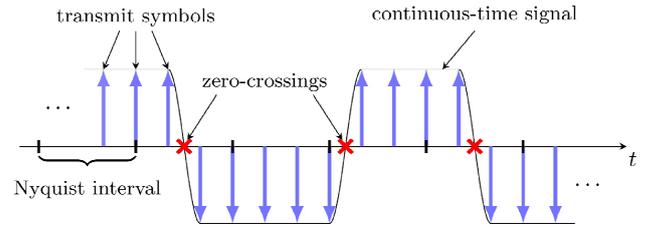

Figure 7: Scheme of the ZXM transmit signal construction for an FTN signaling factor of $M_{Tx} = 3$

If bandwidth is abundant, the trade-off between ADC resolution and bandwidth will be in favor of bandwidth, as the largest possible reduction in ADC resolution while using as much bandwidth as possible is likely optimum in terms of energy-efficiency. In the Gearbox-PHY concept [23], for the case of high spectrum availability it is envisioned to use receivers with 1-bit ADCs in conjunction with temporal oversampling, which are only able to resolve the zero-crossings of the received signal. For such receivers, a natural choice is to encode the information to be conveyed in the distance between the zero-crossings of the transmit signal, a modulation scheme that is termed zero-crossing modulation (ZXM) [24]. A possible way to create a ZXM modulated transmit signal is to use runlength coding in combination with FTN signaling [25]. While FTN signaling introduces intersymbol interference (ISI), the runlength constraint on the transmit symbols governs the minimum distance in which changes in the sign of the transmit symbol can occur and, thus, controls the amount of ISI. A ZXM transmit signal is schematically shown in Figure 7. It is worth noting that although the runlength constraint reduces the amount of information that can be conveyed per symbol, the combination of both runlength coding and FTN signaling increases the information rate to more than 1 bps/Hz per real domain. In [26] runlength encoders and decoders have been derived and spectral efficiencies of up to 4 bps/Hz could be demonstrated. Moreover, for receivers with 1-bit quantization channel estimation and synchronization algorithms have to be re-designed, see, e.g., [27].

An alternative approach to solving the problem of extreme bandwidths is to consider analog multicarrier implementation, in which the ultra-wide band is split into multiple narrower channels and where a dedicated transmit and receive chain is used for each channel, as depicted in Figure 8. The signals are superimposed and upconverted to the corresponding high frequency for a single-stream transmission. Similarly, the receiver employs multiple independent receiver chains corresponding to the narrow channels. This method aims at relaxing the hardware requirements of the transmitter and receiver, primarily the ADC but with possible implications also on the PAs (reduced OOB) and other blocks due to narrowband operation in individual Tx/Rx chains. This architecture provides a high degree of flexibility in spectrum management, as the subcarrier frequency and bandwidth per channel is tunable. For instance, it simplifies the realization of waveforms over disjoint channels, similar to carrier aggregation. Moreover, it allows energy saving by switching off the chains when the full band is not needed. However, in order to improve the spectral efficiency by means of coherent transmission over the subchannels, strict synchronization of the LOs is required.

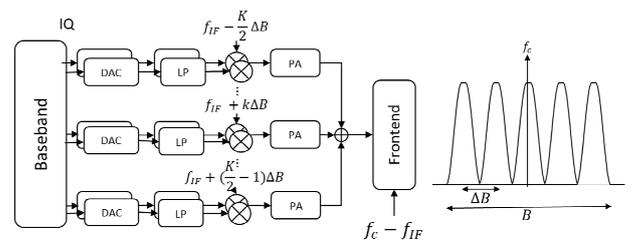

Figure 8: Analog multicarrier transmitter concept

## V. Conclusions

Hardware constraints at sub-THz and ambitious performance requirements for 6G require a rethinking of the waveform and related signal processing. Fulfilling the strict latency requirements needs an order of magnitude decrease of frame processing time. Power amplifier limitations lead to a choice of waveform with small envelope variation. Phase noise in transmitter and receiver requires receiver compensation and a waveform intrinsically resilient to phase noise. Operation at extreme bandwidths requires waveforms that enable the use of ADCs with smaller sampling rate or smaller resolution.

## Acknowledgment

This work has been partly funded by the European Commission through the H2020 project Hexa-X (Grant Agreement no. 101015956).